# Theory of fluid dynamics described from new dynamic hypothesis

Peng SHI[1, 2] *

[1] Logging Technology Research Institute, China National Logging Corporation, No.50 Zhangba five road, Xi'an, 710077, P. R. China

[2] Well Logging Technology Pilot Test Center, China National Logging Corporation, No.50 Zhangba five road, Xi'an, 710077, P. R. China

Corresponding author. Email: sp198911@outlook.com;

ORCID: 0000-0002-8589-5258

**Abstract**. In order to address the difficulties of classical fluid kinematics in describing vorticity and the paradox of linear correlation between viscous force and vorticity in the Navier-Stokes equations, the study examines the inherent relationship between momentum conservation and energy conservation and rederives the fundamental equations of fluid flow from new dynamic hypothesis. The study reveals that momentum conservation and mechanical energy conservation are the same concept termed in different descriptions. The material derivative of velocity is to depict the acceleration of fluid particles in Eulerian perspective. In the study, it is assumed that fluids only obey the theorem of momentum, which describes the translational motion of fluids. The centroid of fluid elements can undergo translational motion under a curl force field. In the hypothesis, the symmetry of stress tensor is determined by the properties of stress field rather than the conservation of moment of momentum for fluid elements. The stress tensor is symmetric when stress field is a curl free field, and the stress tensor is asymmetric when stress field is a curl field. New dynamic equilibrium equations for fluids and the constitutive relationship for Newtonian fluid are proposed under the hypothesis. The Navier-Stokes equations are obtained. The new model of fluid dynamics does not require Stokes hypothesis.

## 1. Introduction

The study on viscous fluid flow is an ancient topic that may date back to prehistoric times when humans began to use streamlined tools like spears and arrows. The attempts to get the exact solution to the problem of the viscous fluid has been made since the Greek philosopher Aristotle (384-322 B.C.) and mathematician Archimedes (287-212 B.C.) [1, 2]. In 1752, Euler first used the partial differential equation to formulate the flow of fluid with negligible viscosity [1, 3]. Subsequently, by adopting Newton's definition of friction, Navier and Stokes independently include the viscous forces into the fluid dynamic equations for inviscid fluids, and the fluid dynamic equations for viscous fluids are obtained, which are now called the Navier-Stokes equations now. Since the Navier-Stokes equations can well describe some special flows, such as laminar pipe flows and some boundary layer flows [1-5], the Navier-Stokes equations are considered the fundamental equations of fluid kinematics. In 2000, seeking the general solution to the Navier-Stokes equations was selected as one of seven Millennium Problems by the Clay Mathematics Institute of Cambridge, U.S. (http://www.claymath.org/millennium-problems). In 2008, the U.S. Defense Advanced Research Projects Agency (DARPA) listed it as one of 23 DARPA Mathematical Challenges— "Mathematical Challenge Four: 21st Century Fluids".

Fluid mechanics is believed to be a branch of classical mechanics. In fluid dynamics, the motion of fluid elements is regarded as the motion of particles, which can be described with Newton's three laws of motion or other mechanical principles related to and equivalent to them [2, 5]. In fluid kinematics, the motion of a fluid is described in two different manners: the Eulerian description and the Lagrangian description. The Navier-Stokes equations are the equations of motion for fluids, in which the viscous force is expressed with velocity. In the Navier-Stokes equations, pressure, velocity, acceleration, etc. are quantities in the Eulerian description. In the description, the momentum of a fluid element is considered a function with respect to time and location coordinates [5]. The acceleration of a fluid element in Eulerian perspective is a combination of local acceleration corresponding to the change of momentum at a point with time and convective acceleration corresponding to the change of momentum at a point with location coordinates. In the theory of fluid dynamics, the conservation of momentum of fluid is described with Cauchy's equation of motion, in which the net force acting on the surface of a free body in fluids is expressed by taking the divergence on the stress tensor, which is a second-order symmetric tensor. In the theory of fluid dynamics, stress is considered to be generated by the deformation of fluid. the relationship between stress deformation of fluid is expressed by the constitutive relationship of fluid, the relationship between stress tensor and the strain rate tensor. Since the stress tensor is a second-order symmetric tensor, the strain rate tensor is a second-order symmetric tensor, which is the symmetric part of the gradient of velocity. The anti-symmetric part of the gradient of velocity is termed as the vorticity tensor. The Navier-Stokes equations are obtained by replacing stress tensor with velocity field through constitutive relationship and geometric relationship.

In the theory of fluid dynamics, the viscous force is generated only by the shear deformation of fluid rather than the shear motion of fluid (or relative slide of fluid elements) defined by Newton. There is a distinction between the viscous forces defined by Newton and the viscous forces defined based on the theory of deformation. The former one believes that viscous force is related to vorticity, while the latter one believes that viscous force is independent of the vorticity. Batchelor appears to have discovered the paradox in the theory of fluid dynamics that the viscous stress can be considered to be generated by deformation or be generated by the vorticity [5]. The paradox is also founded by Liu and Liu (2021) [6]. In their study, they summarized the shortcomings of the theory of fluid dynamics based on the Cauchy-Stokes velocity gradient tensor decomposition, and proposed new governing equations for fluid dynamics by assuming that the viscous stress was generated by the vorticity. The results of several computational examples calculated with the new governing equations for fluid dynamics are compared with that calculated with Navier-Stokes equations, which clearly demonstrate both Navier-Stokes equations and new governing equations have exactly the same results. Considering that the Navier-Stokes equations may work for incompressible and laminar flow but are not satisfied for turbulent flow and compressible flow especially for high-speed flow, new fluid kinematics has been established by Liu in 2021 [7, 8]. Another issue that needs to be pointed out is that during the derivation of Navier-Stokes equations, the stress caused by volume deformation (pressure) has been distinguished from stress tensor. The term describing viscous force in the Navier-Stokes equations should, therefore, not include viscous force related to volume deformation. However, in order to have the thermodynamic pressure equal to the mechanical pressure in fluid, the Stokes hypothesis is introduced in the theory of fluid dynamics. This further suggest that the current fluid dynamics are not fully perfect.

In accordance with Newton's laws of motion, a fluid remains stationary or in a state of uniform linear motion when no external force acts on the fluid, and the motion state of the fluid changes when external forces act on it. The flow of a fluid under the influence of boundaries is the velocity change of fluid particles under boundary resistance. When the Navier-Stokes equations describe the steady flow of an ideal fluid, the equation of mechanical energy conservation, Bernoulli's principle, is derived [5]. Since Bernoulli's principle, the conservation law of mechanical energy for ideal fluids, is obtained from the momentum conservation of ideal fluids, there should be an inherent correlation between momentum conservation and energy conservation, and the inherent correlation should be related to the manner in which the motion of an object is described. In classical mechanics, momentum is the product of a system's mass and its velocity [9]. It is a vector and has the same direction as velocity. Based on Newton's second law of motion, the change in momentum of a system equals the external resultant force multiplied by the time over which its momentum changes. Kinetic energy, a form of energy, is the product of a system's mass and the square of its velocity [9]. It is a scalar. Based on the kinetic energy theorem, the change in kinetic energy equals the external resultant force multiplied by the displacement over which its momentum changes. This indicates that force is described under different perspectives in momentum conservation and energy conservation: force is regarded as a function with respect to time in momentum conservation and a function with respect to position in energy conservation.

The study re-establishes the theory of fluid dynamics from new dynamic hypothesis. The new theory assumes that the theorem of momentum is the only dynamic law that fluids obey, which describes the translational motion of fluids. The centroid of fluid elements can undergo translational motion under a curl force field. New conservation momentum equations for fluid in differential form are proposed. The innate connection between mechanical energy conservation and momentum conservation is examined to further understand the Navier-Stokes equations. It is revealed that the momentum conservation is equivalent to the conservation of mechanical energy.

## 2. Relationship between the properties of a vector field and its gradient

Assuming that $\boldsymbol{A}$ represents an arbitrary vector field, two vectors infinitesimally close in distance satisfy the following connection under linear expansion:

$$\boldsymbol{A}(\boldsymbol{R}+\delta\boldsymbol{R})=\boldsymbol{A}(\boldsymbol{R})+\delta\boldsymbol{R}\cdot\nabla\boldsymbol{A}, \tag{1}$$

where, $\boldsymbol{R}$ is the radius vector and $\delta\boldsymbol{R}$ is the increment of radius vector. Separating the gradient of $\boldsymbol{A}$ into three tensors: a spherical tensor $\boldsymbol{\alpha}$, a deviatoric tensor $\boldsymbol{\alpha}'$ and a rotation tensor $\boldsymbol{\chi}$, they can be expressed with the gradient of $\boldsymbol{A}$ as:

$$\boldsymbol{\alpha}=\frac{1}{6}tr(\nabla\boldsymbol{A}+\boldsymbol{A}\nabla)\boldsymbol{I}, \tag{2}$$

$$\boldsymbol{\alpha}'=\frac{1}{2}(\nabla\boldsymbol{A}+\boldsymbol{A}\nabla)-\boldsymbol{\alpha}, \tag{3}$$

$$\boldsymbol{\chi}=\frac{1}{2}(\nabla\boldsymbol{A}-\boldsymbol{A}\nabla), \tag{4}$$

where, $\boldsymbol{I}$ represents second order unit tensor, $\boldsymbol{A}\nabla$ represents the transposition of $\nabla\boldsymbol{A}$. When vector field $\boldsymbol{A}$ represents velocity field in the theory of fluid dynamics, the spherical tensor $\boldsymbol{\alpha}$, deviatoric tensor $\boldsymbol{\alpha}'$ and rotation tensor $\boldsymbol{\chi}$ correspondingly describe the volume expansion rate, shear deformation rate and rigid body rotation of material element.

With Equations (2) to (4), the divergence and curl of $\boldsymbol{A}$ can be expressed as:

$$\nabla \cdot \boldsymbol{A} = tr\left(\boldsymbol{\alpha} + \boldsymbol{\alpha}' + \boldsymbol{\chi}\right) = tr\left(\boldsymbol{\alpha}\right), \tag{5}$$

$$\nabla \times \boldsymbol{A} = \boldsymbol{\varepsilon} : \left(\boldsymbol{\alpha} + \boldsymbol{\alpha}' + \boldsymbol{\chi}\right) = \boldsymbol{\varepsilon} : \boldsymbol{\chi} . \tag{6}$$

where, $\boldsymbol{\varepsilon}$ is the Levi Civita symbol. Equations (5) and (6) show that the divergence and curl of a vector field are included in spherical tensor and rotation tensor, respectively. When the gradient of $\boldsymbol{A}$ is a symmetric tensor, the vector field $\boldsymbol{A}$ is a curl free field, and when the gradient of $\boldsymbol{A}$ is an asymmetric tensor, the vector field $\boldsymbol{A}$ is a curl field. By taking the divergence of a gradient, the Laplace operator is obtained. The Laplacian of a vector field is considered can be expressed as follows [2]:

$$\nabla^2 \boldsymbol{A} = \nabla \nabla \cdot \boldsymbol{A} - \nabla \times \nabla \times \boldsymbol{A} . \tag{7}$$

Submitting Equations (5) and (6) into Equation (7), the Laplacian of the vector field $\boldsymbol{A}$ can be rewritten as:

$$\nabla^2 \boldsymbol{A} = \nabla\left(tr\left(\boldsymbol{\alpha}\right)\right) - \nabla \times \left(\boldsymbol{\varepsilon} : \boldsymbol{\chi}\right). \tag{8}$$

That is, the Laplacian of a vector field can be described by the spherical tensor and rotation tensor of its gradient.

## 3. Differential description of Newton's second law of motion

When the momentum conservation of a discrete body is described in classical mechanics, the resultant force acting on the discrete body is given by the vector sum.

Figure 1 shows a force system on a block, which is composed of two forces operating vertically on the ends of the block in the $x$ direction. The momentum conservation of the block is expressed by the sum of vectors as:

$$\boldsymbol{f}_1 - \boldsymbol{f}_2 = m\boldsymbol{a}, \tag{9}$$

here, $\boldsymbol{f}_1$ and $\boldsymbol{f}_2$ are the external force acting on both ends of the block, $m$ is the mass of block and $\boldsymbol{a}$ is the average acceleration of the block. Given that the block is a rigid body, the internal force acting on its transversal section along the $x$ direction should rise linearly as illustrated in Figure 1b. The motion of the block under an external force can be described with the motion of any free body included within the block.

Assuming that there is a free body with the length $\triangle x$ located at $x_0$ as illustrated in Figure 1, The momentum conservation of the free body can be expressed by the sum of vectors as:

$$\frac{\boldsymbol{f}_1 - \boldsymbol{f}_2}{L}\Delta x = \frac{\Delta x}{L} m\boldsymbol{a}, \tag{10}$$

here, $L$ is the length of block. The momentum conservation of a free body can be expressed in differential form when its volume approaches toward zero. The momentum conservation of a free body with infinitesimal volume can be expressed in differential form directly using vector operation as:

$$\nabla \cdot \boldsymbol{f}\hat{\boldsymbol{x}} = \rho\boldsymbol{a}, \tag{11}$$

where, $\hat{\boldsymbol{x}}$ is the unit vector of $x$ coordinate and $\rho$ is the density of block. Equation (11) shows that the motion direction should be known beforehand when the momentum conservation of a free body with infinitesimal volume is given in differential form by directly taking the divergence for the internal force field. This is not feasible in continuum mechanics. In order to express the conservation of momentum of a free body with infinitesimal volume in differential form, a new operation must be introduced. This issue is addressed in an attempt by Cauchy's equation of motion. The study believes that the conservation of momentum of fluids can be expressed in other forms.

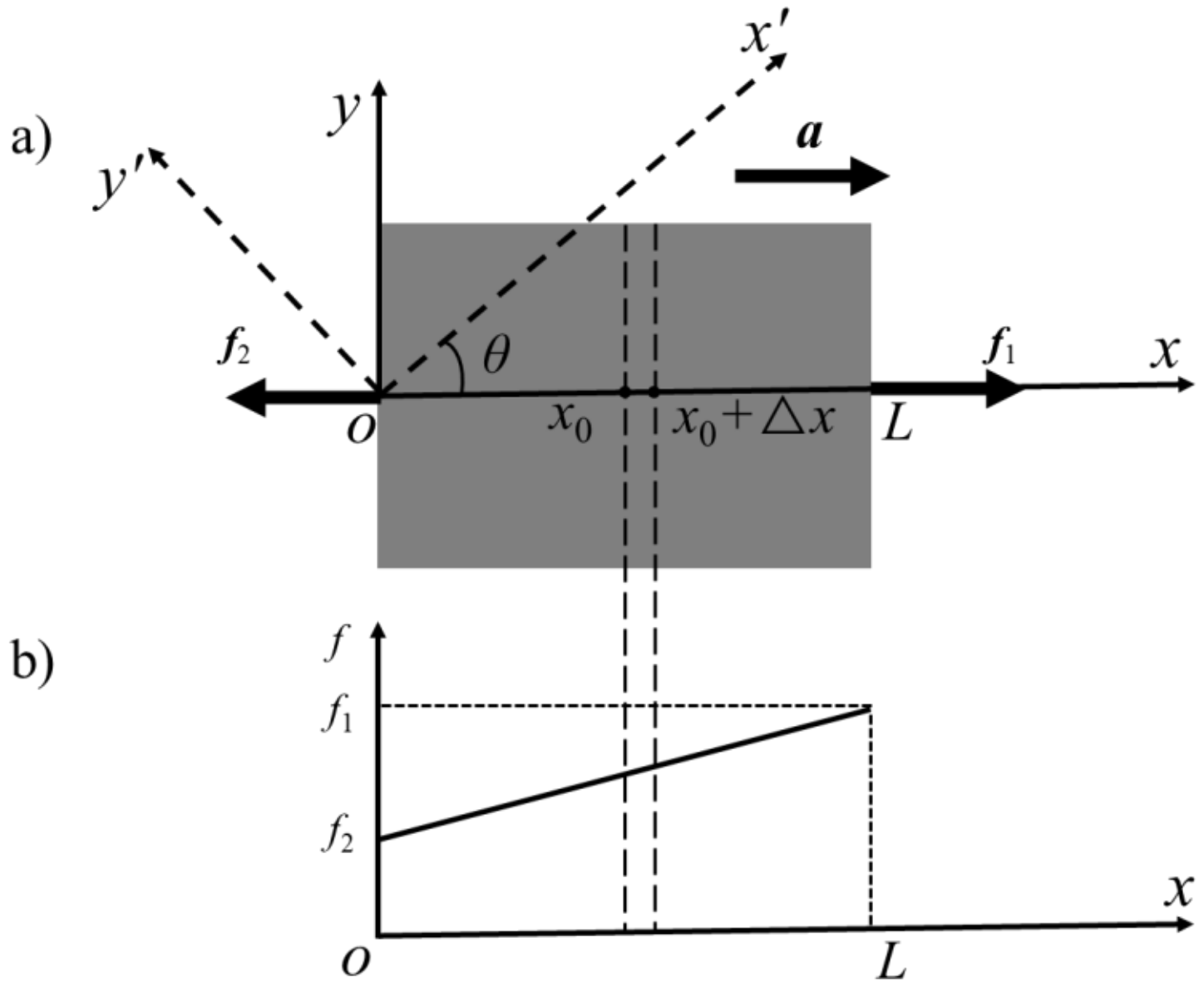

**Figure 1. Force analysis of a block. (a) Forces on the block, (b) internal force on the transversal section of the block along the *x* direction**

For the forced block in Figure 1, the momentum conservation of a free body with infinitesimal volume in the block can be written in differential form as:

$$\nabla \cdot \boldsymbol{\sigma}^S = \rho \boldsymbol{a} \,, \tag{12}$$

here, $\rho$ is the density of block, $\boldsymbol{\sigma}^S$ is a second-order symmetric tensor whose components are expressed with matrix at the Cartesian coordinate system ($x, y, z$) as:

$$\sigma_{ij}^S = \begin{bmatrix} f & 0 & 0 \\ 0 & 0 & 0 \\ 0 & 0 & 0 \end{bmatrix}, \tag{13}$$

with *f* the value of normal force on the transversal section of the block along *x* direction. Since the motion of a block is independent of the selection of coordinates, the momentum conservation of a free body described with Equation (12) should not change with coordinate selection. At the Cartesian coordinate system (*x', y', z'*), the components of $\boldsymbol{\sigma}^S$ are expressed with matrix as:

$$\sigma_{ij}^S = \begin{bmatrix} f\cos^2\theta & f\cos\theta\sin\theta & 0 \\ f\cos\theta\sin\theta & f\sin^2\theta & 0 \\ 0 & 0 & 0 \end{bmatrix}, \tag{14}$$

with $\theta$ the rotation angle of the coordinate system (*x', y', z'*) relative to the coordinate system (*x, y, z*). Given that Equation (12) describes motion irrespective of coordinate choice, the following equation should be true:

$$\nabla\cdot\boldsymbol{\sigma}^S = \nabla\left(tr\left(\boldsymbol{\sigma}^S\right)\right), \tag{15}$$

here, $tr(\boldsymbol{\sigma}^S)$ is the trace of $\boldsymbol{\sigma}^S$, which is the first invariant of $\boldsymbol{\sigma}^S$. It is the obvious thing that a vector can be expanded to a second-order tensor once it is realized that pressure is sometimes considered as a spherical tensor and other times as a scalar.

When the forces illustrated in Figure 1 act on a block, the internal force field in the block is a curl free field because the circulation of surface force acting on any free body inside the block is zero. Figure 2 illustrates a force system on a block, which consists of two forces acting tangentially on the ends of the block in the *x* direction. In contrast to the external forces acting on the block as illustrated in Figure 1, the circulation of surface force acting on the block is non-zero. Assuming that the block is deformable and the acceleration of any free body inside the block is along the direction of resultant force, the conservation of momentum of a free body with infinitesimal volume in the block can be expressed in differential form as:

$$\nabla\times\left(\boldsymbol{\varepsilon}:\boldsymbol{\sigma}^{A}\right)=\rho\boldsymbol{a}, \tag{16}$$

here, $\boldsymbol{\sigma}^{A}$ is a second-order asymmetric tensor whose components can be described by a matrix at the cartesian coordinate system ($x, y, z$) and at the Cartesian coordinate system ($x', y', z'$) respectively as:

$$\sigma_{ij}^{A}=\begin{bmatrix}0 & f & 0\\ 0 & 0 & 0\\ 0 & 0 & 0\end{bmatrix}, \tag{17}$$

and

$$\sigma_{ij}^{A}=\begin{bmatrix}f\cos\theta\sin\theta & f\cos^{2}\theta & 0\\ -f\sin^{2}\theta & -f\cos\theta\sin\theta & 0\\ 0 & 0 & 0\end{bmatrix}. \tag{18}$$

It is easy to prove that $\boldsymbol{\varepsilon}:\boldsymbol{\sigma}^{A}$ is a vector independent of the coordinate selection.

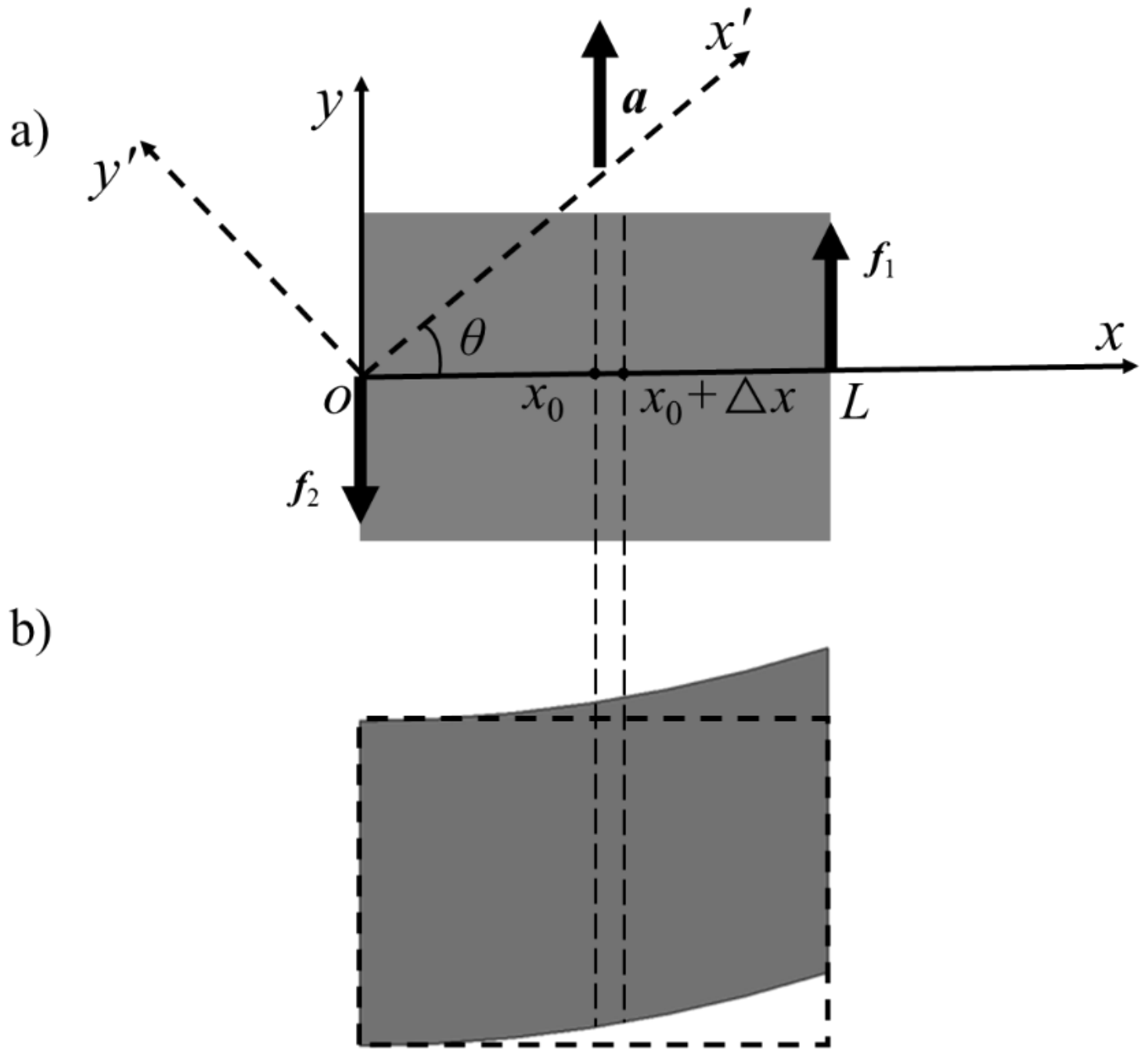

**Figure 2. Force analysis of a block. (a) Forces on the block, (b) motion of the block under external force**

Assuming that the motion of material elements forming a fluid is the superposition of two motions specified above, the conservation of momentum for fluids should be expressed in differential form as follows:

$$\nabla\left(tr\left(\boldsymbol{\sigma}^{S}\right)\right)-\nabla\times\left(\boldsymbol{\varepsilon}:\boldsymbol{\sigma}^{A}\right)=\rho\boldsymbol{a}\,. \tag{19}$$

Comparing Equation (19) with the Navier-Stokes equations, it is seen that they are the same in form [2, 4, 5].

## 4. Correlation between momentum conservation and mechanical energy conservation

Considering the motion of free falling that an object with horizontal velocity is falling under the sole influence of gravity, the trajectory and landing point of the object are determined by its initial velocity as shown in Figure 3. For a motion of free falling with a known initial velocity, the spatial position and time of the object satisfy a one-to-one correspondence during the falling process. Therefore, for any motion of free falling, the velocity and acceleration of the object can be described as a function of its spatial position during its falling process. Under the concept of modern physics, which regards fields as the media of forces, quantities in mechanics like velocity and force are described from the Eulerian perspective as functions with respect to location coordinates.

In Lagrangian perspective, the velocities of a particle at two infinitely close moments satisfy the following relationship:

$$\boldsymbol{v}\left(t+\delta t\right)=\boldsymbol{v}\left(t\right)+\frac{d\boldsymbol{v}}{dt}\delta t\,, \tag{20}$$

here, $\boldsymbol{v}$ is the velocity vector in Lagrangian perspective, $t$ is the time, $\delta t$ is the time increment, and $d/dt$ is the time derivative. In Eulerian perspective, the velocities of a particle at two points that are infinitely close satisfy the following relationship:

$$\boldsymbol{V}\left(\boldsymbol{R}+\delta\boldsymbol{R}\right)=\boldsymbol{V}\left(\boldsymbol{R}\right)+\delta\boldsymbol{R}\cdot\nabla\boldsymbol{V}\ , \tag{21}$$

here, $\boldsymbol{V}$ is the velocity vector in Eulerian perspective which is a function with respect to location coordinates, $\boldsymbol{R}$ is the position vector, $\delta\boldsymbol{R}$ is the displacement of a particle during $\delta t$.

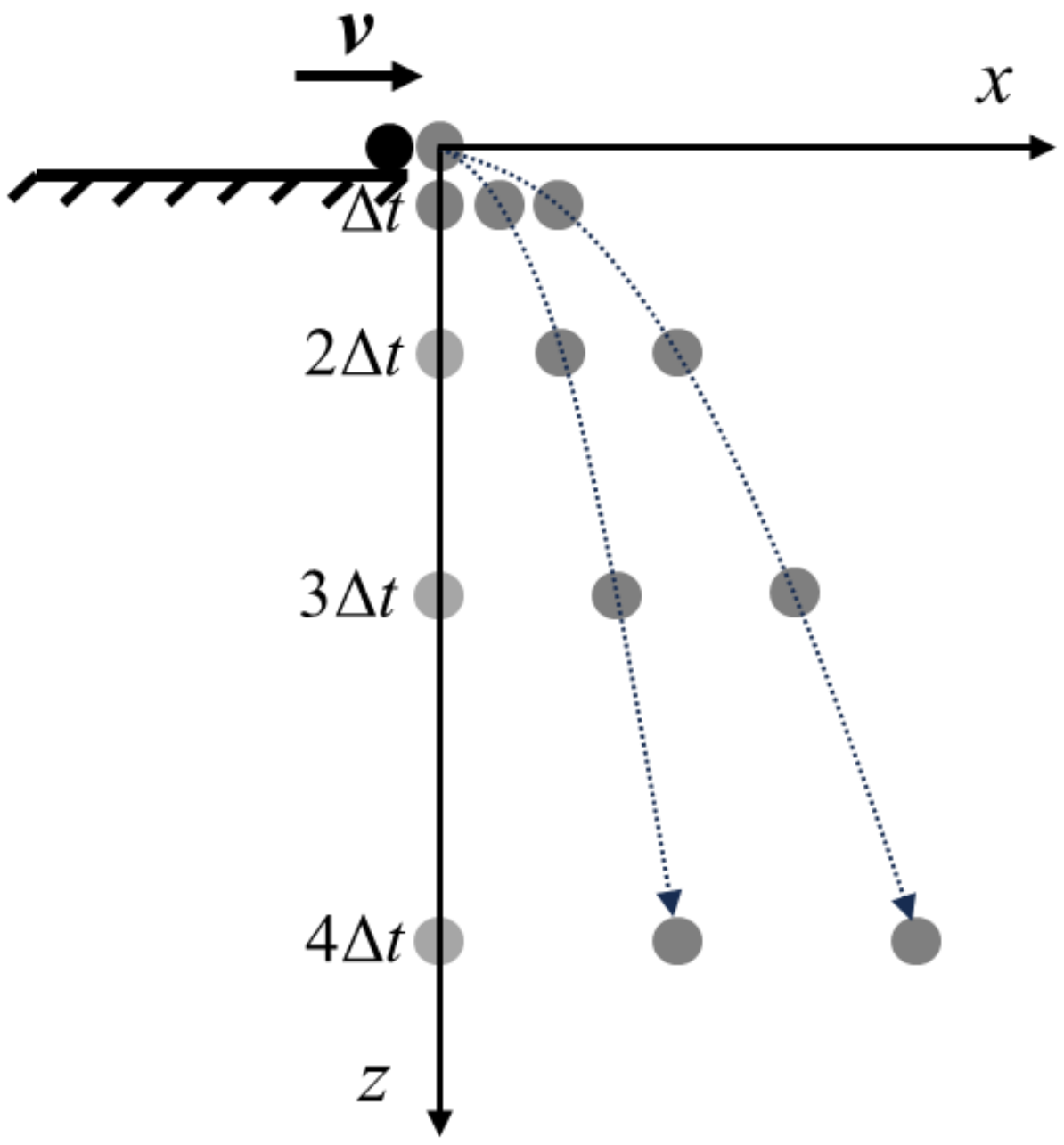

**Figure 3 Diagrams of free fall motion**

Since the following relationship holds at a point:

$$\boldsymbol{V}=\boldsymbol{v}=\frac{\delta\boldsymbol{R}}{\delta t}\ , \tag{22}$$

the velocities described in Lagrangian perspective and in Eulerian perspective meet the following relationship:

$$\frac{d\boldsymbol{v}}{dt} = \boldsymbol{V} \cdot \nabla \boldsymbol{V} . \tag{23}$$

In view of the vector identity [2, 5]

$$\boldsymbol{V} \cdot \nabla \boldsymbol{V} = \frac{1}{2} \nabla \left( V^2 \right) + \nabla \times \boldsymbol{V} \times \boldsymbol{V} , \tag{24}$$

with $V$ the modulus of $\boldsymbol{V}$, Equation (23) is rewritten as:

$$\frac{d\boldsymbol{V}}{dt} = \frac{1}{2} \nabla \left( V^2 \right) + \nabla \times \boldsymbol{V} \times \boldsymbol{V} . \tag{25}$$

For a steady flow, the streamline coincides with the trace, and the velocity at a point in Eulerian perspective is equal to the velocity of a fluid particle at the time that the particle is located at the point in Lagrangian perspective. Hence, the velocity at a streamline can be described as the velocity of a particle moving along the streamline at different times. In accordance with Equation (23), the convective acceleration is the acceleration of a fluid particle described from the Eulerian perspective, by which the acceleration of a fluid particle at a certain moment is obtained from the spatial distribution of velocity in the flow field. Therefore, the Navier-Stokes equations are the equations of motion for fluid in Eulerian perspective, where quantities like velocity, pressure and body force are all field quantities.

In classical mechanics, the following law holds:

$$\boldsymbol{f} \delta t = m \delta \boldsymbol{v} , \tag{26}$$

here, $\boldsymbol{f}$ is the resultant force on an object. Equation (26) is termed the impulse-momentum theorem, by which Newton’s second law is obtained. We know that the external forces described in Lagrangian perspective at a moment are equal to the external forces described in Eulerian perspective at the position where the object is located at this moment. According to the theorem of kinetic energy, the following equation holds:

$$\boldsymbol{F} \cdot \delta \boldsymbol{R} = m\boldsymbol{V} \cdot \nabla \boldsymbol{V} \cdot \delta \boldsymbol{R}, \tag{27}$$

with $\boldsymbol{F}$ the resultant force in Eulerian perspective. Since $\boldsymbol{V}$ is parallel to $\delta\boldsymbol{x}$, which can be obtained from Equation (22), The following equation is obtained:

$$\boldsymbol{V} \times \left(\nabla \times \boldsymbol{V}\right) \cdot \delta \boldsymbol{R} = 0. \tag{28}$$

Then, Equation (27) is rewritten as follows:

$$\boldsymbol{F} \cdot \delta \boldsymbol{R} = \frac{1}{2} m \nabla \left(V^2\right) \cdot \delta \boldsymbol{R}. \tag{29}$$

Equation (29) is the conservation of mechanical energy. Since Equation (29) is obtained from Equation (26) (the equation of momentum conservation), momentum conservation should be equivalent to the conservation of mechanical energy.

According to Newton's second law, the direction of the inertial acceleration of a mass should coincide with the direction of the consequent force exerted upon it. The gradient of a scalar field is a vector field and whose magnitude is the rate of change and which points in the direction of the greatest rate of increase of the scalar field. When the dynamics of a particle are described in field perspective, that is, describing the dynamics of a particle under Eulerian perspective, the gradient direction of kinetic energy is not the direction of the resultant force acting on a particle. Taking Poiseuille flow as example, the gradient direction of kinetic energy field is perpendicular to the streamline of the flow field. Therefore, the second term on the right side of Equation (25) is the velocity change related to the velocity difference between adjacent streamlines at a point, which has nothing to do with the inertial acceleration of fluid particles. This means that the first term on the right side of Equation (25) contains the parts unrelated to particle acceleration, and the second term on the right side of Equation (25) is meant to balance off the parts in the first term. Dou (2006) proposed to character flow instability and turbulence transition by the ratio of the energy gradient in the transverse direction [10]. From the study, it can be concluded that the method predicts flow instability and turbulence transition by comparing the component of the velocity gradient along the streamline to the component of the velocity gradient perpendicular to the streamline.

## 5. Conservation of mass

When the force field is a time-varying field, the acceleration of a particle passing through a specific point in space at different times is different. In this case, the velocities described in Lagrangian perspective and in Eulerian perspective meet the following relationship [2]:

$$\frac{d\boldsymbol{v}}{dt} = \frac{\partial \boldsymbol{V}}{\partial t} + \boldsymbol{V} \cdot \nabla \boldsymbol{V}, \tag{30}$$

which is called the material derivative of velocity. Equation (30) makes clear that, from the Eulerian perspective, the acceleration of fluid particle is made up of two parts: the fluctuation of velocity field over time and the variation of velocity field in space.

Figure 4 shows a moving fluid with a changing density. The fluid density distribution is represented by grayscale in Figure 4. The mass conversation in Eulerian perspective is obtained:

$$\frac{\partial \rho}{\partial t}+\nabla\cdot\left(\rho \boldsymbol{V}\right)=0, \tag{31}$$

with $\rho$ the density of a fluid at a point. When the density of a fluid is uniform and does not change under forces, Equation (31) is simplified as:

$$\nabla\cdot\boldsymbol{V}=0. \tag{32}$$

In this case, mass conversation requires the volume of fluid particle to remain constant. When the fluid moves uniformly, Equation (31) is simplified as:

$$\frac{\partial \rho}{\partial t}+\boldsymbol{V}\cdot\nabla\rho=0. \tag{33}$$

In this case, the density distribution of fluid is what causes the change in density at a given location.

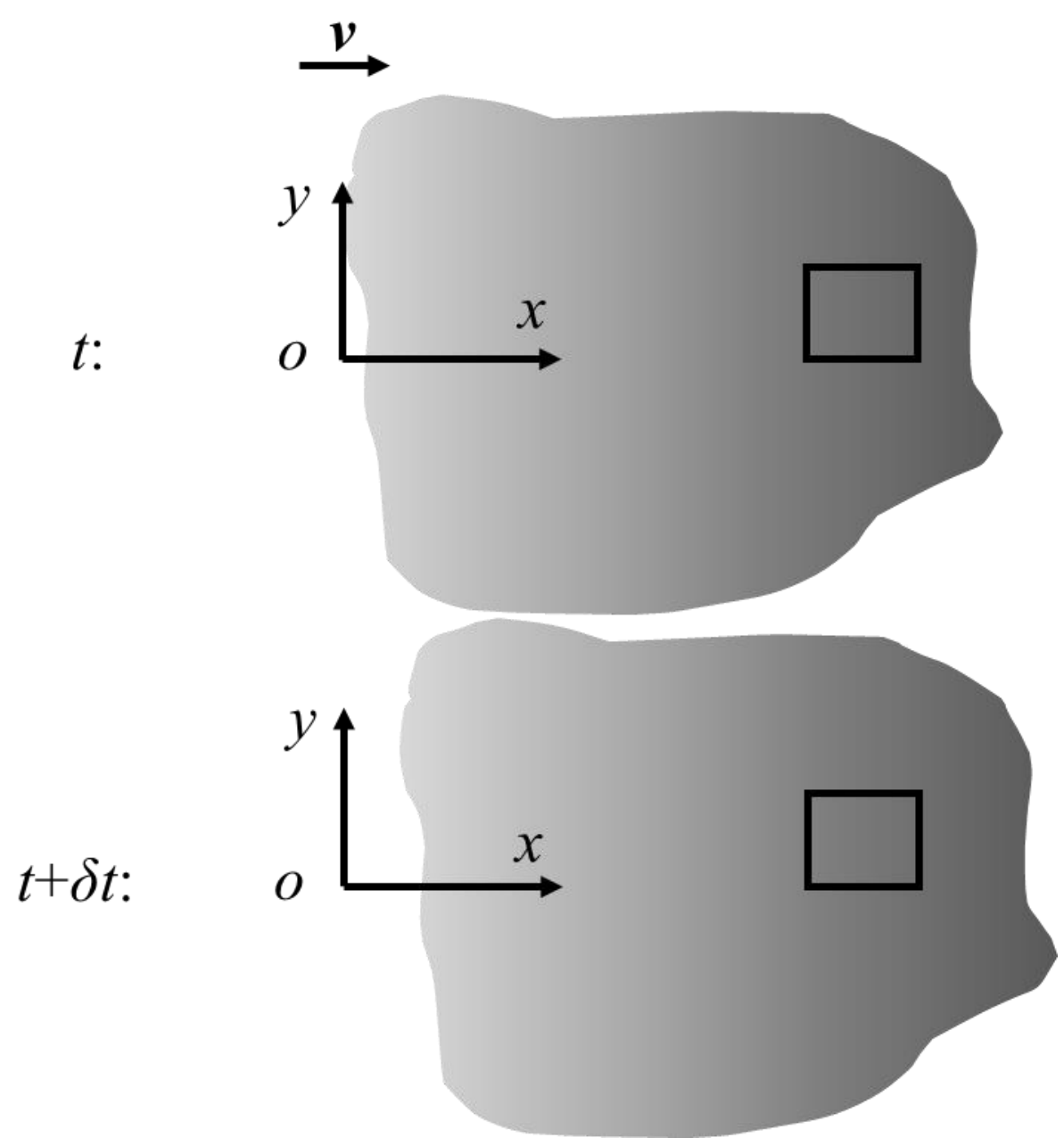

**Figure 4 Diagram of a moving fluid with a changing density**

The momentum of a particle is determined by multiplying its mass by its velocity, hence the change in momentum at a given location should be related with the change in density at that location. In Lagrangian perspective, the rate of momentum change of fluid particles is expressed as follows:

$$\frac{d\left(\bar{\rho}\Omega\mathbf{v}\right)}{dt}=\frac{d\left(\bar{\rho}\Omega\right)}{dt}\mathbf{v}+\bar{\rho}\Omega\frac{d\mathbf{v}}{dt}, \tag{34}$$

here, $\bar{\rho}$ is the density of a fluid particle and $\Omega$ is the volume of the fluid particle. Since the mass conversation of a fluid particle in Lagrangian perspective is expressed as [2]:

$$\frac{d\left(\bar{\rho}\Omega\right)}{dt}=0. \tag{35}$$

Equation (15) is simplified as:

$$\frac{d\left(\bar{\rho}\Omega\mathbf{v}\right)}{dt}=\bar{\rho}\Omega\frac{d\mathbf{v}}{dt}. \tag{36}$$

With the material derivative, the rate of momentum change of a fluid particle with a volume of one unit can be stated in Eulerian perspective as:

$$\frac{d\left(\bar{\rho}\mathbf{v}\right)}{dt}=\rho\frac{\partial\mathbf{V}}{\partial t}+\rho\mathbf{V}\cdot\nabla\mathbf{V}. \tag{37}$$

The term on the right side of Equation (37) is the momentum change of the fluid element described in Eulerian perspective.

## 6. Theory fluid dynamics from new dynamic hypothesis

There is a significant difference between the deformation of fluids and solids under external forces. In fluid dynamics, the direction of the net force is consistent with the flow direction of the fluid. However, in solids, the direction of the net force may not be consistent with the direction of displacement. Therefore, in solids, the stress state at a point cannot be represented by a gradient of a vector, while the stress state at a point in fluids can be represented by a gradient of a vector. According to the analysis of the relationship between a vector field and its gradient in previous section, the stress state at a point in fluids can be expressed with a vector. Symbolling the vector as $\boldsymbol{\Sigma}$ which is termed as the stress vector here, the stress tensor can be written as follows:

$$\boldsymbol{\sigma} = \nabla \boldsymbol{\Sigma} . \tag{38}$$

The characteristic of the stress vector $\boldsymbol{\Sigma}$ controls the circulation of the surface force acting on the free body within a fluid. When the stress vector $\boldsymbol{\Sigma}$ is a curl free field, the circulation of surface force acting on the free body is zero and the stress tensor is a symmetric tensor. When the stress vector $\boldsymbol{\Sigma}$ is a curl field, the circulation of surface force acting on the free body is non-zero and the stress tensor is an asymmetric tensor. With Equation (38), the divergence of the stress tensor can be rewritten with $\boldsymbol{\Sigma}$ as:

$$\nabla \cdot \boldsymbol{\sigma} = \nabla \cdot \nabla \boldsymbol{\Sigma} . \tag{39}$$

In accordance of Equation (8), Equation (39) can be rewritten as:

$$\nabla \cdot \boldsymbol{\sigma} = \nabla \left( tr\left( \boldsymbol{\sigma}^S \right) \right) - \nabla \times \left( \boldsymbol{\varepsilon} : \boldsymbol{\tau} \right), \tag{40}$$

where, $\boldsymbol{\sigma}^S$ is the symmetric part of $\boldsymbol{\sigma}$, and $\boldsymbol{\tau}$ is the antisymmetric part of $\boldsymbol{\sigma}$, which are respectively expressed with $\boldsymbol{\Sigma}$ as:

$$\boldsymbol{\sigma}^S = \frac{1}{2}\left( \nabla \boldsymbol{\Sigma} + \boldsymbol{\Sigma} \nabla \right), \tag{41}$$

$$\boldsymbol{\tau} = \frac{1}{2}\left( \nabla \boldsymbol{\Sigma} - \boldsymbol{\Sigma} \nabla \right). \tag{42}$$

For viscous fluids, the stress is related to the volume deformation and shear motion. When the shear motion occurs in a viscous fluid, the velocity field is a curl field. In this case, the internal force field in the viscous fluid is a curl field, and the stress tensor is an asymmetric tensor. According to the characteristics of fluid flow, the curl force field and the curl free force field are coupled. When the stress brought on by shear motion is distinguished from the stress brought on by volume deformation, the momentum conservation of fluid is expressed as follows:

$$\nabla\left(tr\left(\boldsymbol{\sigma}\right)\right)-\nabla\times\left(\boldsymbol{\varepsilon}:\boldsymbol{d}\right)+\boldsymbol{F}=\rho\frac{\mathrm{D}\boldsymbol{V}}{\mathrm{D}t}, \tag{43}$$

with

$$\boldsymbol{d}=\boldsymbol{\sigma}-\frac{1}{3}tr\left(\boldsymbol{\sigma}\right)\boldsymbol{I}, \tag{44}$$

here, $\boldsymbol{d}$ is the deviatoric stress tensor which is an asymmetric tensor. For a Newtonian fluid, the relation between deviatoric stress tensor and deviatoric strain rate tensor $\boldsymbol{\xi}$ in component form is expressed as:

$$d_{ij}=\mu\delta_{ik}\delta_{jl}\xi_{kl}, \tag{45}$$

where, $\mu$ is the viscosity of fluid. The deviatoric strain rate tensor $\boldsymbol{\xi}$ and the velocity $\boldsymbol{V}$ have the following relationship:

$$\boldsymbol{\xi}=\nabla\boldsymbol{V}-\frac{1}{3}tr\left(\nabla\boldsymbol{V}\right)\boldsymbol{I}. \tag{46}$$

Submitting Equation (46) into Equation (44) and consequently submitting the relation between deviatoric stress and deviatoric strain rate expressed with velocity into Equation (43), the momentum conservation of a fluid can be expressed with velocity as:

$$\nabla\left(tr\left(\boldsymbol{\sigma}\right)\right)-\mu\nabla\times\left(\boldsymbol{\varepsilon}:\nabla\boldsymbol{V}\right)+\boldsymbol{F}=\rho\frac{\mathrm{D}\boldsymbol{V}}{\mathrm{D}t}. \tag{47}$$

Since the following relations hold:

$$\nabla\times\left(\boldsymbol{\varepsilon}:\nabla\boldsymbol{V}\right)=\nabla\times\nabla\times\boldsymbol{V}=-\nabla^{2}\boldsymbol{V}+\nabla\nabla\cdot\boldsymbol{V}, \tag{48}$$

Equation (47) can be rewritten as:

$$\nabla\left(tr\left(\boldsymbol{\sigma}\right)\right)+\mu\left(\nabla^2\boldsymbol{V}-\nabla\nabla\cdot\boldsymbol{V}\right)+\boldsymbol{F}=\rho\frac{\mathrm{D}\boldsymbol{V}}{\mathrm{D}t}. \tag{49}$$

In accordance with the definition of pressure, the following equation holds:

$$p=-\frac{1}{3}tr\left(\boldsymbol{\sigma}\right), \tag{50}$$

with $p$ defined pressure. Equation (49) is rewritten as:

$$-3\nabla p+\mu\left(\nabla^2\boldsymbol{V}-\nabla\nabla\cdot\boldsymbol{V}\right)+\boldsymbol{F}=\rho\frac{\mathrm{D}\boldsymbol{V}}{\mathrm{D}t}. \tag{51}$$

Equation (51), while different, has the same in form as the Navier-Stokes equations for Newtonian fluid. The difference between Equation (51) and the Navier-Stokes equations indicates that there is a difference between regarding pressure as a scalar and a spherical tensor.

When pressure is regarded as a scalar symbolled $P$, the relationship between fluid volume deformation and pressure is expressed as:

$$P=-K\nabla\cdot\boldsymbol{U}, \tag{52}$$

with $K$ the bulk modulus of fluid and $\boldsymbol{U}$ is the displacement field. When pressure is regarded as a spherical tensor, the relationship between fluid volume deformation and pressure is expressed as:

$$p\boldsymbol{I}=-\frac{K}{3}\left(\nabla\cdot\boldsymbol{U}\right)\boldsymbol{I}. \tag{53}$$

It is seen from Equations (52) and (53) that $P$=3$p$. The pressure in the Navier-Stokes equations is the pressure in scalar form $P$.

For further special case of incompressible fluid, the mass conservation reduces to $\nabla\cdot\boldsymbol{V}=0$, which means that the density of fluid does not change with pressure,

Equation (51) reduces to:

$$-\nabla P + \mu\nabla^2\boldsymbol{V} + \boldsymbol{F} = \rho\frac{\mathrm{D}\boldsymbol{V}}{\mathrm{D}t}. \tag{54}$$

Equation (54) is the equation of motion for incompressible Newtonian fluid.

From Equation (47), it is seen that the viscous force is generated only by the shear motion of fluid (or relative slide of fluid elements) rather than shear deformation of fluid. As a result, Newton's definition of the viscous force differs significantly from the one based on deformation theory. There are two key distinctions between the viscous forces as defined by Newton and the viscous forces based on deformation theory. One is that the former one implies that the potential flow is an inviscid flow, while the latter one implies that the potential flow is a viscous flow. Another difference is that the former one believes that viscous force is related to vorticity, while the latter one believes that viscous force is independent of vorticity. In the Navier-Stokes equations, the viscous force is only related to the curl velocity field. This illustrates that the viscous force defined by Newton is more appropriate than that defined based on deformation theory.

When the fluid flow is described with the theory of fluid dynamics based on traditional deformation theory, the bulk viscosity is added in the viscous coefficient tensor, which indicates that the viscous force can be caused by bulk deformation of fluid [2]. In order to make the thermodynamic pressure equal to the mechanical pressure in fluid, the Stokes hypothesis is introduced. In fact, during the formulation of Navier-Stokes equations, the stress caused by volume deformation (or pressure) has been distinguished from stress tensor. The term describing viscous force in the Navier-Stokes equations should, therefore, not include viscous force related to volume deformation. This again suggests that the model of traditional fluid dynamics has problems in explaining the fluid dynamics.

## 7. Conclusions

In summary, the study rederived the fundamental equations of fluid dynamics from new dynamic hypothesis and investigated the intrinsic relationship between momentum conservation and energy conservation. The study assumes that the theorem of momentum is the only dynamic law that fluids obey. The centroid of fluid elements undergoes translational motion under a curl force field. Under the hypothesis, new dynamic equilibrium equations for fluid and the constitutive relationship for Newtonian fluid are proposed. Then, the Navier-Stokes equations are obtained. Under the new model of fluid dynamics, the Stokes hypothesis is no longer required. The symmetry of stress tensor is determined by the properties of stress field rather than the conservation of moment of momentum for fluid elements. The stress tensor is symmetric when stress field is a curl free field, and the stress tensor is asymmetric when stress field is a curl field. It has been discovered that momentum conservation and mechanical energy conservation are two different terms for the same concept, and the material derivative of velocity field is to describe the acceleration of fluid particles in Eulerian perspective. The gradient direction of kinetic energy is not the direction of the resultant force applied on a particle when the dynamics of a particle are described from the Eulerian perspective.

**Acknowledgments**

This work was supported by the scientific research and technology development project of China National Petroleum Corporation (2020B-3713) and the scientific research and technology development project of China National Logging Corporation (F-D60023KA).

**Competing Interests**

The authors have no conflicts to disclose.

**References**

[1] Anderson, J.: "A History of Aerodynamics, and Its Impact on Flight Machines", Cambridge Aerospace Series, vol. 8. Cambridge University Press, Cambridge (1997).

[2] White, F.M., Corfield, I.: Viscous fluid flow. McGraw-Hill, New York (2006).

[3] Sheng, W.: A revisit of Navier–Stokes equation. *European Journal of Mechanics B-fluids*, 80: 60-71 (2020).

[4] Schlichting, H., Gersten, K.: Boundary-layer theory. Springer, Heidelberg (2016).

[5] Batchelor, G.K.: An introduction to fluid dynamics. Cambridge University Press, Cambridge (2000).

[6] Liu, C., Liu Z.: New governing equations for fluid dynamics. *AIP Advances*, 11, 115025 (2021).

[7] Liu, C.: New fluid kinematics. *Journal of Hydrodynamics*, 33(3), 395-399 (2021).

[8] Liu, C.: New ideas on governing equations of fluid dynamics. *Journal of Hydrodynamics*, 33(4): 861-866 (2021).

[9] Kibble, T.W.B., Berkshire, F.H.: Classical mechanics, 5th edition. Imperial College Press, London (2004).

[10] Dou, H.: Mechanism of flow instability and transition to turbulence. *International Journal of Non-Linear Mechanics*, 41, 512–517 (2006).